\documentclass[12pt]{amsart}
\usepackage{amssymb,latexsym,amsmath,enumitem,mathrsfs,geometry}
\usepackage{thmtools}
\usepackage{fullpage}
\usepackage{fancyhdr}
\usepackage{xcolor}
\usepackage{bbm}
\usepackage{stmaryrd}
\usepackage{mathrsfs}
\usepackage{todonotes}
\usepackage{hyperref}
\makeatletter
\renewcommand{\enddoc@text}{%
  \ifx\@empty\@translators\else
    \@settranslators
  \fi
  \ifx\@empty\addresses\else
    \clearpage
    \phantomsection
    \addcontentsline{toc}{section}{Authors' Affiliations}
    \@setaddresses
  \fi
}
\usepackage{eso-pic}
\usepackage{graphicx}
\usepackage{subcaption}
\usepackage{diagbox}
\usepackage{array}
\usepackage{tikz}
\usetikzlibrary{arrows}
\usetikzlibrary{positioning, arrows.meta, shapes.geometric}
\usepackage{float}
\usepackage{comment}
\usepackage{mathtools}
\usepackage{amsthm}
\usepackage{placeins}
\usepackage{acro}
\usepackage{setspace}
\usepackage{lineno}
\newcommand{\N}{\mathcal{N}}
\newcommand{\R}{\mathbb{R}}
\newcommand{\E}{\mathbb{E}}

\newcommand{\Beta}{\operatorname{Beta}}
\newcommand{\KL}{\operatorname{KL}}
\newcommand{\vol}{\operatorname{vol}}
\newcommand{\diag}{\operatorname{diag}}
\newtheorem{theorem}{Theorem}[section]
\newtheorem{definition}[theorem]{Definition}
\newtheorem{corollary}[theorem]{Corollary}
\newtheorem{proposition}[theorem]{Proposition}
\newtheorem{lemma}[theorem]{Lemma}
\newtheorem{remark}[theorem]{Remark}

\DeclareAcronym{RV}{
short = RV,
long  = \emph{Random Variable},
tag   = abbrev}
\DeclareAcronym{KL}{
short = KL,
long  = \emph{Kullback--Leibler},
tag   = abbrev}
\DeclareAcronym{GM}{
short = GM,
long  = \emph{Gaussian Mixture},
tag   = abbrev}
\DeclareAcronym{PDF}{
short = PDF,
long  = \emph{Probability Density Function},
tag   = abbrev}
\DeclareAcronym{EM}{
short = EM,
long  = \emph{Expectation-Maximization},
tag   = abbrev}
\usepackage{cleveref}
\usepackage[
    backend=biber,
    style=numeric,
    sorting=none,
    doi=true,
    url=true,
    eprint=true,
    isbn=true
]{biblatex}
\begin{document}
\numberwithin{equation}{section}
\title{
    Closed-Form Gaussian Information Projection for Embedded Hypercubes and Beta--Gaussian Mixtures
    }
\author{Ali Darijani\(^{\circledast\square\#\P\perp}\)}
\address{
Ali Darijani\(^{\circledast\square\#\P\perp}\)\newline
\indent ORCID: \url{https://orcid.org/0009-0000-4659-093X}\newline
\indent Research Staff, Fraunhofer IOSB of Fraunhofer-Gesellschaft, Germany\newline
\indent Research Staff, KIT Department of Computer Science, Germany\newline
\indent Alumnus, RWTH Aachen University, Germany\newline
\indent Alumnus, Sharif University of Technology, Iran
}
\email{ali.darijani@iosb.fraunhofer.de}
\email{ali.darijani@kit.edu}
\email{ali.darijani@rwth-aachen.de}
\email{ali.darijani@alum.sharif.edu}
\author{Emadedin Soufian\(^{\square\#\P\perp}\)}
\address{
Emadedin Soufian\(^{\square\#\P\perp}\)\newline
\indent ORCID: \url{https://orcid.org/0009-0006-5777-4623}\newline
\indent Research Staff, Det Norske Veritas (DNV), Germany\newline
\indent Alumnus, RWTH Aachen University, Germany\newline
\indent Alumnus, Iran University of Science and Technology, Iran
}
\email{emadedin.soufian@dnv.com}
\email{emadedin.soufian@rwth-aachen.de}
\author{Jürgen Beyerer\(^{\circledast\perp}\)}
\address{
Jürgen Beyerer\(^{\circledast\perp}\)\newline
\indent ORCID: \url{https://orcid.org/0000-0003-3556-7181}\newline
\indent Director, Fraunhofer IOSB of Fraunhofer-Gesellschaft, Germany\newline
\indent Professor, KIT Department of Computer Science, Germany\newline
\indent Alumnus, Karlsruhe Institute of Technology, Germany
}
\email{juergen.beyerer@iosb.fraunhofer.de}
\email{juergen.beyerer@kit.edu}
\address{
Symbol Legend\newline
\indent\({\circledast}\): Idea, \({\square}\): Math, \({\#}\): Programming, \({\P}\): Writing, \({\perp}\): Critique
}

\thanks{Funded by the Deutsche Forschungsgemeinschaft (DFG) - Project number 471687386}
\date{\today}
\begin{abstract}
We derive closed-form Gaussian information projections of 
Beta--Gaussian distributions whose tangential components are 
supported on affine embedded hypercubes and whose normal 
components are Gaussian. The optimal Gaussian is uniquely 
characterized by moment matching, and the corresponding 
minimum Kullback--Leibler divergence admits an explicit 
expression depending only on the intrinsic Beta parameters, 
independently of the embedding geometry, ambient dimension, 
and normal Gaussian parameters. In the uniform case, this 
divergence reduces to a linear function of the intrinsic 
dimension. We further extend this construction to finite, 
pure embedded cubical complexes by considering mixtures of 
the componentwise Gaussian projections. Using the joint convexity 
of the Kullback--Leibler divergence, we establish an explicit 
upper bound on the approximation error of the resulting Gaussian Mixture. 
For uniform tangential distributions and volume-based mixture weights, 
the bound is linear in the intrinsic dimension and independent of 
the geometry of the complex, its embedding, and the normal 
Gaussian parameters.
\end{abstract}
\maketitle
\tableofcontents
\AddToShipoutPictureBG{
    \AtPageCenter{
        \begin{tikzpicture}[remember picture,overlay]
            \node[
                rotate=-45,
                text=gray!25
            ] at (current page.center) {
                \upshape\normalfont\resizebox{0.9\paperwidth}{!}{\bfseries Preprint}
            };
        \end{tikzpicture}
    }
}
\section{Introduction}
\label{sec:introduction}
Gaussian \acp{PDF} provide a tractable representation 
of probability measures, but their approximation quality 
can deteriorate when the underlying \acp{PDF} exhibit 
bounded or geometrically constrained tangential structure. 
This motivates the study of Gaussian information projections for 
\acp{PDF} associated with embedded geometric structures.

In this work, we consider affine embeddings of the unit hypercube 
into an ambient Euclidean space. We equip the reference hypercube with 
product-Beta \acp{PDF} and introduce independent Gaussian \acp{PDF} 
in the orthogonal complement of the embedding's tangent space. 
This construction gives absolutely continuous \acp{PDF} with respect to 
ambient Lebesgue measure, while preserving the prescribed Beta \ac{PDF} 
structure in the tangential directions. We refer to these \acp{PDF} as 
Beta--Gaussian \acp{PDF}.

Our main result provides a closed-form characterization of the unique 
Gaussian information projection of a Beta--Gaussian \acp{PDF} 
through moment matching. Moreover, the resulting minimum \ac{KL} 
divergence admits an explicit expression depending only on the intrinsic 
Beta \ac{PDF} parameters, independently of the embedding 
geometry, ambient dimension, and normal Gaussian parameters. 
In the uniform case, the divergence reduces to a linear function 
of the intrinsic dimension.

We further extend this construction to finite, pure embedded cubical complexes 
by considering mixtures of the componentwise Beta--Gaussian \acp{PDF} 
and their corresponding Gaussian information projections. Using the 
joint convexity of the \ac{KL} divergence, we establish an 
explicit upper bound for the approximation error of the resulting \ac{GM}. 
In the uniform case, with volume-based mixture weights, this 
bound is again linear in the intrinsic dimension and independent of 
the geometry of the complex, its embedding, and the normal Gaussian \ac{PDF} 
parameters.
\section{Construction}
\label{sec:construction}

\begin{definition}[Convex Set~\cite{convex_anal}]
Let \(d\in\mathbb{N}\). A set \(K\subseteq\mathbb{R}^d\) is
called convex if, for every \(x,y\in K\) and every
\(t\in[0,1]\), the convex combination
\begin{equation}
    (1-t)x+ty
\end{equation}
belongs to \(K\).
\end{definition}

\begin{definition}[Standard Cube~\cite{cube}]
For \(d\in\mathbb{N}\), the standard \(d\)-dimensional cube is
the set
\begin{equation}
    Q_d\coloneq[0,1]^d =\{x\in\mathbb{R}^d:0\leq x_i\leq 1\text{ for }i=1,\ldots,d\}.
\end{equation}

\end{definition}

\begin{definition}[Embedded Cube]
Let \(1\leq d\leq D\), let \(v\in\mathbb{R}^D\), and let
\(E\in\mathbb{R}^{D\times d}\) have full column rank, that is,
\begin{equation}
    \operatorname{rank}(E)=d.
\end{equation}
The affine map
\begin{equation}
    \Phi:\mathbb{R}^d\to\mathbb{R}^D,\quad \Phi(x)\coloneq v+Ex,
\end{equation}
is injective. The image of the standard \(d\)-dimensional cube
under \(\Phi\),
\begin{equation}
    C:=\Phi(Q_d)
    =\{v+Ex:x\in Q_d\},
\end{equation}
is called an embedded \(d\)-dimensional cube in
\(\mathbb{R}^D\).
\end{definition}

\begin{definition}[Tangent and Normal Spaces]
Let \(C=\Phi(Q_d)\subseteq\mathbb{R}^D\) be an embedded
\(d\)-dimensional cube, with
\begin{equation}
    \Phi(x)=v+Ex,\quad v\in\mathbb{R}^D,\quad E\in\mathbb{R}^{D\times d},\quad\operatorname{rank}(E)=d.
\end{equation}
We equip \(\mathbb{R}^D\) with its standard Euclidean inner
product
\begin{equation}
    \langle x,y\rangle\coloneq x^\top y,\quad x,y\in\mathbb{R}^D.
\end{equation}
The tangent space of \(C\), understood as the linear direction
space of its affine hull, is defined by
\begin{equation}
    T_C\coloneq\operatorname{range}(E)=\{Eh:h\in\mathbb{R}^d\}\subseteq\mathbb{R}^D.
\end{equation}
Equivalently, since \(\Phi\) is affine, its differential is
constant and satisfies
\begin{equation}
    D\Phi(x)=E
    \qquad\text{for every }x\in\mathbb{R}^d,
\end{equation}
so that
\begin{equation}
    T_C=\operatorname{range}(D\Phi(x))\quad\text{for every }x\in\mathbb{R}^d.
\end{equation}
The normal space of \(C\), relative to the ambient Euclidean
space \(\mathbb{R}^D\), is defined as the orthogonal complement
of its tangent space:
\begin{equation}
    N_C\coloneq T_C^\perp=\left\{z\in\mathbb{R}^D:\langle z,w\rangle=0\text{ for every }w\in T_C\right\}.
\end{equation}
Equivalently, the normal space is the null space of the
transpose of \(E\):
\begin{equation}
    N_C=\ker(E^\top)=\left\{z\in\mathbb{R}^D:E^\top z=0\right\}.
\end{equation}
In particular, the ambient space admits the orthogonal direct
sum decomposition
\begin{equation}
    \mathbb{R}^D=T_C\oplus N_C,\quad T_C\perp N_C,
\end{equation}
with dimensions
\begin{equation}
    \dim(T_C)=d,\quad\dim(N_C)=D-d.
\end{equation}
\end{definition}

\begin{remark}[Geometric Interpretation]
The embedded cube \(C\) is a \(d\)-dimensional parallelotope
contained in the affine subspace
\begin{equation}
    \operatorname{aff}(C)=v+\operatorname{range}(E)=v+T_C.
\end{equation}
Its \(d\) edge vectors are precisely the columns of \(E\).
In particular, the columns of \(E\) are linearly independent,
but they are not necessarily mutually orthogonal or of equal
length. Consequently, the term embedded cube refers to the
affine image of the standard cube and does not imply that
\(C\) is an orthogonal cube in the ambient Euclidean metric.
The translation vector \(v\) determines its position, whereas
\(E\) determines its orientation, edge directions, and edge
lengths.
\end{remark}

\begin{proposition}
    Let \(C=F_C([0,1]^d)\), where \(F_C(x)=v+Dx\) and
    \(D\in\mathbb{R}^{n\times d}\) has full column rank. Then
    \begin{equation}
        \operatorname{vol}_d(C)=\sqrt{\det(D^\top D)}.
        \label{eq:cube-volume}
    \end{equation}
\end{proposition}

\begin{proof}
    By the area formula (see \cite[Theorem~3.2.3,~p.~243]{federer_gmt}),
    \begin{equation}
      \operatorname{vol}_d(C)=\int_{[0,1]^d}\sqrt{\det(D^\top D)}dx=\sqrt{\det(D^\top D)}.
    \end{equation}
\end{proof}

\begin{definition}[Faces of an Affine Cube]
Let \(C=v+E[0,1]^d\) be an embedded affine \(d\)-cube.
For \(I\subseteq\{1,\ldots,d\}\) and
\(\epsilon\in\{0,1\}^I\), define
\begin{equation}
F_{I,\epsilon}(C)\coloneq v+E\bigl\{x\in[0,1]^d : x_i=\epsilon_i \text{ for all } i\in I\bigr\}.
\end{equation}
These sets are the faces of \(C\), with
\begin{equation}
    \dim F_{I,\epsilon}(C)=d-|I|.
\end{equation}
\end{definition}

\begin{definition}[Embedded Cubical Complex]
A finite collection \(\mathcal K\) of embedded affine cubes is an embedded cubical complex 
if
\begin{enumerate}
    \item \(F\in\mathcal K\) whenever \(F\) is a face of
    \(C\in\mathcal K\);
    \item \(C\cap C'\) is either empty or a face of both
    \(C,C'\in\mathcal K\).
\end{enumerate}
The complex is pure of dimension \(d\) if
\begin{equation}
    \dim C=d\quad\text{for every inclusion-maximal } C\in\mathcal K.
\end{equation}
Its realization and maximal cells are denoted by
\begin{equation}
    |\mathcal K|\coloneq\bigcup_{C\in\mathcal K}C,\quad\mathcal K^{[d]}\coloneq\{C\in\mathcal K:\ C\text{ is inclusion-maximal}\}.
\end{equation}
\end{definition}

\begin{proposition}[Cubical Volume Decomposition]
\label{prop:cubical-volume}
Let \(\mathcal K\) be a finite, pure, \(d\)-dimensional embedded cubical complex in \(\mathbb R^D\). Then
\begin{equation}
    \mathcal H^d(|\mathcal K|)=\sum_{C\in\mathcal K^{[d]}}\vol_d(C).
\end{equation}
Moreover, if \(0<\mathcal H^d(|\mathcal K|)<\infty\), then
\begin{equation}
    \mathcal U_{|\mathcal K|}=\sum_{C\in\mathcal K^{[d]}}\frac{\vol_d(C)}{\mathcal H^d(|\mathcal K|)}\mathcal U_C,\quad\mathcal U_A\coloneq\frac{\mathcal H^d|_A}{\vol_d(A)}.
\end{equation}
\end{proposition}

\begin{proof}
Distinct maximal cells intersect in proper faces, which are
$\mathcal H^d$-null by \cite[Lemma~2.2,~p.~92]{evans_measure}.
Finite additivity of $\mathcal H^d$ then follows from
\cite[Theorem~2.1.3,~p.~54]{federer_gmt}.
\end{proof}

\begin{definition}[Product-Beta \ac{PDF}~\cite{beta_handbook}]
\label{def:product-beta}
For parameter vectors \(a=(a_1,\ldots,a_d)\in(0,\infty)^d\) and 
\(b=(b_1,\ldots,b_d)\in(0,\infty)^d\), let \(X=(X_1,\ldots,X_d)^\top\) 
have independent coordinates
\begin{equation}
  X_i\sim\Beta(a_i,b_i),\quad i=1,\ldots,d.
\end{equation}
Its \ac{PDF} with respect to Lebesgue measure on \([0,1]^d\) is
\begin{equation}
  f_{a,b}(x)=\prod_{i=1}^d\frac{x_i^{a_i-1}(1-x_i)^{b_i-1}}{B(a_i,b_i)},\quad x\in(0,1)^d,
  \label{eq:product-beta-density}
\end{equation}
where \(B(a,b)=\Gamma(a)\Gamma(b)/\Gamma(a+b)\) is the beta function~\cite[\S~5.12,~p.~142]{func_hand}. 
The \ac{PDF} is set to zero outside \([0,1]^d\).
\end{definition}

\begin{proposition}[Moments and Entropy of a Product-Beta Law]
\label{prop:beta-moments-entropy}
Under \cref{def:product-beta}, define, for each \(i\),
\begin{equation}
  \mu_i=\frac{a_i}{a_i+b_i},\quad s_i^2=\frac{a_i b_i}{(a_i+b_i)^2(a_i+b_i+1)}.
  \label{eq:beta-moments}
\end{equation}
Then
\begin{equation}
  \mathbb{E}(X)=\mu,\quad \mathbb{V}(X)=V\coloneq\diag(s_1^2,\ldots,s_d^2)\succ0.
  \label{eq:product-beta-moments}
\end{equation}
The differential entropy is finite and equals
\begin{equation}
  h(X)=\sum_{i=1}^d h_i,
  \label{eq:product-beta-entropy}
\end{equation}
where
\begin{equation}
 h_i={}\log B(a_i,b_i)-(a_i-1)\psi(a_i)-(b_i-1)\psi(b_i)\notag+(a_i+b_i-2)\psi(a_i+b_i),
 \label{eq:beta-entropy}
\end{equation}
and \(\psi\) is the digamma function.
\end{proposition}

\begin{proof}
The moment identities are the standard univariate 
Beta moment formulas, together with independence. 
For entropy, taking the expectation of the logarithm 
of the \ac{PDF} gives
\begin{equation}
  -h(X_i)=\E(\log f_{X_i}(X_i))=-\log B(a_i,b_i)+(a_i-1)\E[\log X_i]+(b_i-1)\E(\log(1-X_i)).
\end{equation}
Differentiating \(\log B(a,b)\) with respect to its two arguments, or equivalently differentiating the beta integral under the integral sign, gives
\begin{equation}
  \E(\log X_i)=\psi(a_i)-\psi(a_i+b_i),\quad\E(\log(1-X_i))=\psi(b_i)-\psi(a_i+b_i).
\end{equation}
Substitution gives \cref{eq:beta-entropy}. These expectations are finite for all strictly positive parameters: near either endpoint, \(x^{a_i-1}|\log x|\) and \((1-x)^{b_i-1}|\log(1-x)|\) are integrable. Thus each marginal entropy is finite. Independence gives additivity of entropy and \cref{eq:product-beta-entropy}.
\end{proof}

\begin{definition}[Beta Law on an Embedded Hypercube]
\label{def:embedded-beta}
Let \(C=v+E[0,1]^d\) be an embedded hypercube. 
For \(a,b\in(0,\infty)^d\), the embedded product-Beta 
law on \(C\) is the pushforward of the product-Beta 
law in \cref{def:product-beta} under \(F_C(x)=v+Ex\). In other words,
\begin{equation}
  Y_T=v+EX,\qquad X_i\stackrel{\mathrm{ind}}{\sim}\Beta(a_i,b_i).
  \label{eq:embedded-beta-rv}
\end{equation}
We denote its law by \(P_T(C;a,b)\). Its intrinsic \ac{PDF} is 
understood with respect to \(\mathcal H^d\) on the affine hull of \(C\).
\end{definition}

\begin{proposition}[Embedded Moments and Intrinsic Entropy]
\label{prop:embedded-beta-moments}
Under \cref{def:embedded-beta},
\begin{equation}
  m_T\coloneq\E(Y_T)=v+E\mu,\quad \Sigma_T\coloneq\mathbb{V}(Y_T)=EVE^\top.
  \label{eq:embedded-beta-mean-var}
\end{equation}
The covariance is positive definite on \(T_C\) and zero on \(N_C\). The intrinsic differential entropy is
\begin{equation}
  h_{T_C}(Y_T)=\sum_{i=1}^d h_i+\frac12\log\det(E^\top E).
  \label{eq:embedded-beta-entropy}
\end{equation}
\end{proposition}

\begin{proof}
The moment formulas follow from the affine transformation rule. 
Since \(E\) has full column rank and \(V\succ0\), the restriction 
of \(EVE^\top\) to \(T_C=\operatorname{range}(E)\) is positive definite. 
For the entropy, the constant \(d\)-dimensional Jacobian 
of \(F_C\) is \(J_C=\sqrt{\det(E^\top E)}\). The change-of-variables 
formula on the affine hull gives \(h_{T_C}(Y_T)=h(X)+\log J_C\), which 
is \cref{eq:embedded-beta-entropy}.
\end{proof}

\begin{definition}[Beta--Gaussian \ac{PDF}]
\label{def:beta-gaussian}
Let \(C=v+E[0,1]^d\subseteq\R^D\) be an embedded hypercube, with \(d\leq D\). 
Choose matrices \(U_T\in\R^{D\times d}\) and \(U_N\in\R^{D\times(D-d)}\) 
whose columns are orthonormal bases of \(T_C\) and \(N_C\), respectively. Let
\begin{equation}
  X_i\stackrel{\mathrm{ind}}{\sim}\Beta(a_i,b_i),\quad Z_N\sim\N_{D-d}(\nu_N,S_N),\qquad S_N\succ0,
\end{equation}
with \(X\) and \(Z_N\) independent. Define
\begin{equation}
  W\coloneq v+EX+U_NZ_N.
  \label{eq:beta-gaussian-rv}
\end{equation}
The law of \(W\) is called a Beta--Gaussian distribution associated 
with \(C\), and is denoted by \(P(C;a,b,\nu_N,S_N)\). If \(d=D\), 
the normal term is absent and all statements involving \(U_N,Z_N,S_N\) 
are interpreted on the zero-dimensional space.
\end{definition}

\begin{lemma}[Absolute Continuity and Finite Entropy]
\label{lem:beta-gaussian-entropy}
The Beta--Gaussian \ac{RV} in \cref{def:beta-gaussian} is absolutely continuous 
with respect to \(D\)-dimensional Lebesgue measure and has 
finite differential entropy. More precisely, for \(d<D\),
\begin{equation}
 h(W)=\sum_{i=1}^d h_i+\frac12\log\det(E^\top E)+\frac{D-d}{2}\log(2\pi e)+\frac12\log\det S_N.
 \label{eq:beta-gaussian-entropy}
\end{equation}
\end{lemma}

\begin{proof}
The product-Beta vector has a \ac{PDF} on \((0,1)^d\) and finite 
entropy by \cref{prop:beta-moments-entropy}. Its affine image has finite 
intrinsic entropy by \cref{prop:embedded-beta-moments}. The Gaussian normal 
coordinate has a nondegenerate \ac{PDF} and finite entropy. Under the 
orthogonal affine coordinate map
\begin{equation}
  \Phi(w)=
  \begin{pmatrix}
    U_T^\top(w-v)\\ 
    U_N^\top(w-v)
  \end{pmatrix},
\end{equation}
the absolute Jacobian determinant is one and
\begin{equation}
  \Phi(W)=
  \begin{pmatrix}
    U_T^\top EX
    \\ Z_N
  \end{pmatrix}.
\end{equation}
The two blocks are independent. Their product law is absolutely 
continuous in \(\R^D\), and entropy additivity gives 
\cref{eq:beta-gaussian-entropy}. The same argument without 
the normal block applies when \(d=D\).
\end{proof}

\begin{remark}
The normal Gaussian is essential when \(d<D\): without it 
the distribution is supported on a \(d\)-dimensional 
affine subspace and is singular with respect to 
ambient Lebesgue measure. The positive-definite normal 
covariance makes the ambient density nondegenerate. No 
isotropy assumption is imposed on \(S_N\).
\end{remark}

\begin{definition}[Forward \ac{KL} divergence]
For Lebesgue \acp{PDF} \(p,q\) on \(\R^D\), define
\begin{equation}
  \KL(P:Q)=\int_{\R^D}p(x)\log\frac{p(x)}{q(x)}\mathrm{d}x,
\end{equation}
with the usual extended-real convention. 
Let \(\mathcal G_D\) denote the family of nondegenerate 
Gaussian \acp{PDF} on \(\R^D\). The information projection 
of \(P\) onto \(\mathcal G_D\) is any minimizer of 
\(\KL(P:Q)\) over \(Q\in\mathcal G_D\).
\end{definition}

\begin{proposition}[Gaussian Information Projection by Moment Matching]
\label{prop:moment-projection}
Let \(P\) be absolutely continuous on \(\R^D\), have finite 
differential entropy, finite second moments, and positive-definite 
covariance \(\Sigma_P\). Then its unique information projection 
onto \(\mathcal G_D\) is
\begin{equation}
  Q^*=\N_D(\cdot\mid m_P,\Sigma_P),
  \qquad m_P=\E_P[X].
  \label{eq:moment-projection}
\end{equation}
\end{proposition}

\begin{proof}
For \(Q=\N_D(\cdot\mid m,\Sigma)\), direct integration gives
\begin{equation}
 \KL(P:Q)={}-h(P)+\frac D2\log(2\pi)+\frac12\log\det\Sigma\notag+\frac12\operatorname{tr}(\Sigma^{-1}\Sigma_P)+\frac12(m-m_P)^\top\Sigma^{-1}(m-m_P).
 \label{eq:kl-gaussian-objective}
\end{equation}
For fixed \(\Sigma\succ0\), the last term is uniquely minimized at 
\(m=m_P\). To minimize over \(\Sigma\), set 
\(A=\Sigma_P^{-1/2}\Sigma\Sigma_P^{-1/2}\). Up to constants 
independent of \(A\), the objective is \(\log\det A+\operatorname{tr}(A^{-1})\). 
If \(\lambda_1,\ldots,\lambda_D\) are the positive eigenvalues of 
\(A\), this is \(\sum_k(\log\lambda_k+\lambda_k^{-1})\), whose 
unique minimum occurs at \(\lambda_k=1\) for every 
\(k\) (the scalar function has derivative \((\lambda-1)/\lambda^2\) 
and a unique global minimum at one). Thus \(\Sigma=\Sigma_P\), 
proving existence and uniqueness.
\end{proof}

\begin{theorem}[Closed-form Projection of a Beta--Gaussian Law]
\label{thm:beta-gaussian-projection}
Let \(P=P(C;a,b,\nu_N,S_N)\) be the Beta--Gaussian law in \cref{def:beta-gaussian}. 
Define \(\mu,V\) by \cref{eq:beta-moments}, and let 
\(\overline\Sigma_N=U_NS_NU_N^\top\) (zero if \(d=D\)). 
Then the unique Gaussian information projection of \(P\) onto 
\(\mathcal G_D\) is
\begin{equation}
  Q^*=\N_D(\cdot\mid m,\Sigma),
  \label{eq:beta-gaussian-projection}
\end{equation}
where
\begin{equation}
  m=v+E\mu+U_N\nu_N,\quad \Sigma=EVE^\top+\overline\Sigma_N.\label{eq:beta-gaussian-projection-m-var}
\end{equation}
Its minimum divergence is
\begin{equation}
 \KL(P:Q^*)=C_d(a,b):=\sum_{i=1}^d\left[\frac12\log(2\pi e s_i^2)-h_i\right],
 \label{eq:beta-gaussian-constant}
\end{equation}
where \(s_i^2\) and \(h_i\) are given by \cref{eq:beta-moments,eq:beta-entropy}. 
In particular, the minimum divergence is independent of the ambient dimension, 
the embedding \(v,E\), the hypercube volume, the normal mean, and the normal covariance.
\end{theorem}

\begin{proof}
Independence of \(X\) and \(Z_N\) gives the mean and covariance 
in \cref{eq:beta-gaussian-projection-m-var}. The tangential 
covariance is positive definite on \(T_C\), while \(\overline\Sigma_N\) 
is positive definite on \(N_C\); since these spaces are orthogonal complements, 
\(\Sigma\succ0\). The entropy is finite by \cref{lem:beta-gaussian-entropy}, 
and the second moments are finite because the Beta component is bounded 
and the normal component is Gaussian. Hence \cref{prop:moment-projection} 
applies and proves the stated unique projection.

It remains to compute the divergence. In orthonormal coordinates adapted 
to \(T_C\oplus N_C\), both \(P\) and \(Q^*\) factor into a tangential 
and a normal distribution. Their normal factors agree exactly, so 
invariance and additivity of KL divergence give
\begin{equation}
  \KL(P:Q^*)=\KL(P_T:Q_T^*).
  \label{eq:normal-cancellation}
\end{equation}
The entropy of the tangential target is, by \cref{eq:embedded-beta-entropy},
\begin{equation}
 h_{T_C}(Y_T)=\sum_{i=1}^d h_i+\frac12\log\det(E^\top E).
\end{equation}
Let \(U_T\) be an orthonormal basis matrix for \(T_C\) and put \(R=U_T^\top E\). 
Since \(E=U_TR\) and \(R\) is invertible, the matrix of the tangential 
covariance in this orthonormal basis is \(RVR^\top\). Therefore
\begin{equation}
  \det_{T_C}(\Sigma_T)=\det(R)^2\det(V)\notag=\det(E^\top E)\prod_{i=1}^d s_i^2.
  \label{eq:tangent-determinant}
\end{equation}
The entropy of the moment-matched tangential Gaussian is consequently
\begin{equation}
 h(Q_T^*)=\frac d2\log(2\pi e)+\frac12\log\det(E^\top E)+\frac12\sum_{i=1}^d\log s_i^2.
\end{equation}
Subtracting the target entropy and using \cref{eq:normal-cancellation} 
gives \cref{eq:beta-gaussian-constant}. All geometric 
and normal terms cancel, as claimed.
\end{proof}

\begin{corollary}[Uniform \ac{PDF} on an Embedded Hypercube]
\label{cor:uniform-cube}
If \(a_i=b_i=1\) for all \(i\), then \(X\) is uniform on \([0,1]^d\), \(\mu_i=1/2\), \(s_i^2=1/12\), and \(h_i=0\). Hence
\begin{equation}
  m=v+\frac12E\mathbf{1}_d+U_N\nu_N,\quad\Sigma=\frac1{12}EE^\top+\overline\Sigma_N,
\end{equation}
and
\begin{equation}
  \KL(P:Q^*)=\frac d2\log\left(\frac{\pi e}{6}\right).
  \label{eq:uniform-cube-constant}
\end{equation}
The divergence depends only on the intrinsic dimension \(d\).
\end{corollary}

\begin{proof}
For the uniform Beta\((1,1)\) law, the stated moments and entropy follow immediately 
from the univariate formulas. Substitution into \cref{thm:beta-gaussian-projection} 
gives the result.
\end{proof}

\begin{remark}[Nonnegativity]
Each summand in \cref{eq:beta-gaussian-constant} is the \ac{KL} divergence 
between a univariate Beta\((a_i,b_i)\) density and its moment-matched Gaussian on 
\(\R\). It is therefore nonnegative. Equality cannot occur for a 
nondegenerate Beta law, since its support is bounded whereas a 
nondegenerate Gaussian has unbounded support. Thus \(C_d(a,b)>0\) for \(d\geq1\).
\end{remark}

\begin{definition}[Finite \ac{GM}]
A finite \ac{GM} on \(\R^D\) is a \ac{PDF}
\begin{equation}
 q(x)=\sum_{j=1}^M\lambda_j\N_D(x\mid m_j,\Sigma_j),\quad \lambda_j\geq0,\quad \sum_{j=1}^M\lambda_j=1,\quad \Sigma_j\succ0.
\end{equation}
\end{definition}

\begin{proposition}[{Joint Convexity of \ac{KL} divergence~\parencite[Proposition~2.14,~p.~92]{info_geo}}]
\label{prop:kl-joint-convexity}

Let \(p_j,q_j\) be \acp{PDF} with respect to a common measure and let 
\(\lambda_j\geq0\) with \(\sum_j\lambda_j=1\). Then
\begin{equation}
 \KL\left(\sum_{j=1}^M\lambda_jp_j:\sum_{j=1}^M\lambda_jq_j\right)
 \leq\sum_{j=1}^M\lambda_j\KL(p_j:q_j),
 \label{eq:joint-convexity}
\end{equation}
where the inequality is understood in the extended-real sense.
\end{proposition}

\begin{proof}
For almost every \(x\), apply the log-sum inequality~\parencite[Lemma~2.10,~p.~93]{info_geo}
\begin{equation}
 \left(\sum_j r_j\right)\log\frac{\sum_jr_j}{\sum_js_j}
 \leq\sum_jr_j\log\frac{r_j}{s_j}
\end{equation}
with \(r_j=\lambda_jp_j(x)\) and \(s_j=\lambda_jq_j(x)\), using the standard zero conventions. 
Integration gives \cref{eq:joint-convexity}.
\end{proof}

\begin{theorem}[Beta--Gaussian Mixture Approximation on an Embedded Cubical Complex]
\label{thm:beta-gaussian-mixture}
Let \(\mathcal K\) be a finite, pure, \(d\)-dimensional embedded cubical 
complex in \(\R^D\), with maximal cells \(C_1,\ldots,C_M\). For 
each \(j\), write
\begin{equation}
 C_j=v_j+E_j[0,1]^d,
\end{equation}
where \(E_j\) has full column rank. Choose 
parameters \(a_j,b_j\in(0,\infty)^d\), a normal mean \(\nu_{N,j}\), 
and a positive-definite normal covariance \(S_{N,j}\) on \(N_{C_j}\) (with 
the normal term omitted if \(d=D\)). Let \(P_j\) be the corresponding 
Beta--Gaussian law and let \(q_j\) be its unique Gaussian 
information projection as specified by \cref{thm:beta-gaussian-projection}. 
For arbitrary mixture weights
\begin{equation}
 \lambda_j>0,\qquad \sum_{j=1}^M\lambda_j=1,
\end{equation}
define
\begin{equation}
 P=\sum_{j=1}^M\lambda_jP_j,\quad Q=\sum_{j=1}^M\lambda_jq_j.
 \label{eq:beta-gaussian-mixtures}
\end{equation}
Then each \(q_j\) is the unique Gaussian information projection of \(P_j\), and
\begin{equation}
 \KL(P:Q)\leq\sum_{j=1}^M\lambda_j C_d(a_j,b_j),
 \label{eq:beta-gaussian-mixture-bound}
\end{equation}
where \(C_d\) is given in \cref{eq:beta-gaussian-constant}. In particular, 
the bound is independent of the ambient dimension, the embeddings of the 
cells, and all normal means and covariance matrices.
\end{theorem}

\begin{proof}
The componentwise information-projection statement and the identity \(\KL(P_j:q_j)=C_d(a_j,b_j)\) follow from \cref{thm:beta-gaussian-projection}. Applying joint convexity, \cref{prop:kl-joint-convexity}, to the mixtures in \cref{eq:beta-gaussian-mixtures} gives
\begin{equation}
 \KL(P:Q)\leq\sum_{j=1}^M\lambda_j\KL(P_j:q_j)=\sum_{j=1}^M\lambda_j C_d(a_j,b_j),
\end{equation}
which proves the bound. Its stated invariances follow term by 
term from \cref{thm:beta-gaussian-projection}.
\end{proof}

\begin{corollary}[Uniform Cubical--Complex Approximation]
\label{cor:uniform-cubical-complex}
Under the assumptions of \cref{thm:beta-gaussian-mixture}, 
suppose \(0<\mathcal H^d(|\mathcal K|)<\infty\). On each maximal 
cell let the tangential law be uniform, i.e. \(a_{j,i}=b_{j,i}=1\) for 
all \(i,j\), and choose the volume weights
\begin{equation}
 \lambda_j=\frac{\vol_d(C_j)}{\mathcal H^d(|\mathcal K|)}.
\end{equation}
Let \(P=\sum_j\lambda_jP_j\) be the resulting mixture after 
independent normal Gaussian smoothing on each cell, and let 
\(Q=\sum_j\lambda_jq_j\) be the mixture of the componentwise 
Gaussian information projections. Then
\begin{equation}
 \KL(P:Q)\leq\frac d2\log\left(\frac{\pi e}{6}\right).
 \label{eq:uniform-cubical-bound}
\end{equation}
The bound depends only on the intrinsic dimension \(d\) and is 
independent of the ambient dimension, the geometry and 
embedding of the cubical complex, and the normal 
Gaussian parameters.
\end{corollary}

\begin{proof}
By \cref{prop:cubical-volume}, the volume weights are positive 
and sum to one. By \cref{cor:uniform-cube}, every component divergence 
equals \(\frac d2\log(\pi e/6)\). Substitution into \cref{thm:beta-gaussian-mixture} 
proves \cref{eq:uniform-cubical-bound}.
\end{proof}

\begin{remark}[What the Mixture Bound Does and Does not Assert]
The mixture in \cref{thm:beta-gaussian-mixture} is obtained by projecting each component separately and retaining the original mixture weights. The theorem proves a constructive upper bound for this particular \ac{GM}. It does not assert that \(Q\) globally minimizes \(\KL(P:\widetilde Q)\) over all \acp{GM} \(\widetilde Q\), nor that the displayed upper bound is the optimal value of such a global mixture optimization problem.
\end{remark}

\begin{remark}[Uniform Measure Versus its Smoothed Version]
The volume-weighted identity is an identity of probability 
measures with respect to \(d\)-dimensional Hausdorff measure 
on the cubical complex. If \(d<D\), this unsmoothed measure is 
singular with respect to \(D\)-dimensional Lebesgue measure. The \ac{KL} 
guarantee in \cref{cor:uniform-cubical-complex} concerns the 
mixture after adding the componentwise normal Gaussian perturbations, 
which produces ambient Lebesgue densities; it is not a finite 
ambient-density KL statement for the original singular uniform measure.
\end{remark}
\section{Conclusion}

We established a closed-form characterization of the forward Gaussian information 
projection of Beta--Gaussian \acp{PDF} associated with affine embedded hypercubes. 
The optimal Gaussian \ac{PDF} is uniquely determined by moment matching, 
and the corresponding minimum \ac{KL} divergence admits an explicit expression 
depending only on the intrinsic Beta parameters, independently of the embedding 
geometry, ambient dimension, and normal Gaussian \ac{PDF} parameters. In the 
uniform \ac{PDF} case, this divergence reduces to a linear function of the 
intrinsic dimension.

We further extended this construction to finite, pure embedded cubical 
complexes by constructing \acp{GM} from componentwise information projections. 
Joint convexity of the \ac{KL} divergence gives an explicit upper bound on 
the representation divergence of the resulting \ac{GM}. For uniform 
tangential \acp{PDF} and volume-based mixture weights, the bound reduces 
to a linear function of the intrinsic dimension, independently of the geometry 
and embedding of the cubical complex and the normal Gaussian \ac{PDF} parameters.

These results provide a dimension-independent framework for quantifying Gaussian \ac{PDF} 
and \ac{GM} representation divergence of \acp{PDF} associated with embedded 
geometric structures.
\section{Acknowledgment}
This work was funded by Project SFB~1574/CRC~1574 ``A Circular Factory 
for the Perpetual Product'', Deutsche Forschungsgemeinschaft (DFG), 
Project ID 471687386.
\printbibliography
\end{document}